\documentclass[twocolumn,superscriptaddress,aps,prl]{revtex4}
\usepackage{graphicx}

\begin{document}
\bibliographystyle{prsty}
\title{Ultrafast pump-probe study of phase separation and competing orders in the underdoped (Ba,K)Fe$_{2}$As$_{2}$
superconductor}
\author{Elbert E. M. Chia}
\affiliation{Division of Physics and Applied Physics, School of
Physical and Mathematical Sciences, Nanyang Technological
University, Singapore 637371, Singapore}
\author{D. Talbayev}
\affiliation{Los Alamos National Laboratory, Los Alamos, NM 87545,
USA}
\author{Jian-Xin Zhu}
\affiliation{Los Alamos National Laboratory, Los Alamos, NM 87545,
USA}
\author{H. Q. Yuan}
\affiliation{Department of Physics, Zhejiang University, Hangzhou,
Zhejiang 310027, China} \affiliation{Los Alamos National Laboratory,
Los Alamos, NM 87545, USA}
\author{T. Park}
\affiliation{Los Alamos National Laboratory,
Los Alamos, NM 87545, USA} \affiliation{Department of Physics,
Sungkyunkwan University, Suwon 440-746, South Korea}
\author{J. D. Thompson}
\affiliation{Los Alamos National Laboratory, Los Alamos, NM 87545,
USA}
\author{C. Panagopoulos}
\affiliation{Division of Physics and Applied Physics, School of
Physical and Mathematical Sciences, Nanyang Technological
University, Singapore 637371, Singapore}
\affiliation{Department of Physics, University of Crete and
FORTH, 71003 Heraklion, Greece}
\author{G. F. Chen}
\affiliation{Beijing National Laboratory for Condensed Matter
Physics, Institute of Physics, Chinese Academy of Sciences, Beijing
100080, China}
\author{J. L. Luo}
\affiliation{Beijing National Laboratory for Condensed Matter
Physics, Institute of Physics, Chinese Academy of Sciences, Beijing
100080, China}
\author{N. L. Wang}
\affiliation{Beijing National Laboratory for Condensed Matter
Physics, Institute of Physics, Chinese Academy of Sciences, Beijing
100080, China}
\author{A. J. Taylor} \affiliation{Los Alamos National
Laboratory, Los Alamos, NM 87545, USA}
\date{\today}

\begin{abstract}
We report measurements of quasiparticle relaxation dynamics in
the high-temperature superconductor (Ba,K)Fe$_{2}$As$_{2}$ in
optimally doped, underdoped and undoped regimes. In the
underdoped sample, spin-density wave (SDW) order forms at
$\sim$85~K, followed by superconductivity at $\sim$ 28~K. We
find the emergence of a normal-state order that suppresses SDW
at a temperature $T^{\ast} \sim$ 60~K and argue that this
normal-state order is a precursor to superconductivity.
\end{abstract}

\maketitle
\begin{figure} \centering \includegraphics[width=8cm,clip]{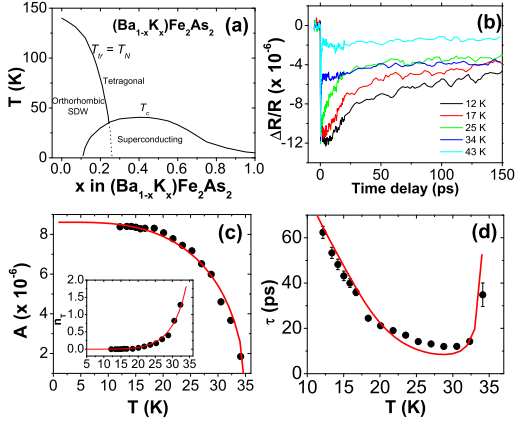}
\caption{(a) Phase diagram of BKFA,
with $T_{c}$ and structural phase transition temperature ($T_{tr}$)
(adapted from \cite{Rotter08b}). $T_{N}$
coincides with $T_{tr}$ until both are suppressed \cite{ChenH09}.
(b) $\Delta R/R$ versus pump-probe delay of
optimally doped BKFA ($T_{c} \sim$ 36~K).
(c) Amplitude $A$ extracted from single-exponential decay fits. Inset: Density
of thermally-excited QPs, $n_{T}(T)$. Solid lines
are fits to the RT model, yielding $\Delta(0) =
3.0k_{B}T_{c}$. (d) Relaxation time
$\tau(T)$. Solid line is the fit to the RT model.}
\label{fig:Fig1}
\end{figure}

 The recently discovered FeAs-based pnictides
\cite{Kamihara08,ChenXH08,Wu08,Ni08,Rotter08} constitute the
only class of superconductors (SCs), besides the cuprates, with
superconducting transition temperatures ($T_{c}$) exceeding
50~K. The pnictides have a layered structure like the cuprates,
with FeAs planes instead of CuO$_{2}$ planes. Like many other
SCs on the border of magnetism, such as the organics
\cite{Moser98}, heavy-fermions \cite{Mathur98}, and cuprate
high-temperature superconductors (HTSCs)
\cite{Kivelson03,Sachdev03,Hufner08}, the pnictides exhibit a
rich phase diagram, with antiferromagnetism (or spin density
wave, SDW) at low dopings \cite{Cruz08} and SC at intermediate
dopings. Figure~\ref{fig:Fig1}(a) shows the phase diagram of a
particular family of pnictides --- (Ba,K)Fe$_{2}$As$_{2}$
(BKFA) \cite{Rotter08b}, which is the subject of study in this
Letter. These phases in BKFA are mesoscopically separated in
the underdoped compound, with SDW and SC/normal state regions
\cite{Park09}. Moreover, inelastic neutron scattering in an
optimally-doped BKFA revealed the presence of a 14~meV magnetic
resonance mode in the SC phase, localized in both energy and
wavevector \cite{Christianson08}. A large Fe-isotope effect was
seen in BKFA, suggesting the role played by magnetic
fluctuations in superconductivity \cite{Liu09}. In all these
classes of SCs, how these phases interact with one another, and
the role of magnetism, are open questions that might help
understand superconductivity in these compounds.

In the cuprate HTSCs, femtosecond pump-probe spectroscopy has
shown to be a useful tool to discern coexisting/competing
phases, such as the pseudogap phase in
YBa$_{2}$Cu$_{2}$O$_{7-x}$ (Y-123) \cite{Kabanov99}, and the
suppression of superconductivity by antiferromagnetism in the
tri-layered Tl$_{2}$Ba$_{2}$Ca$_{2}$Cu$_{3}$O$_{y}$
\cite{Chia07}. In this Letter, we use the pump-probe technique
to study the BKFA SCs in the undoped, underdoped and
optimally-doped regimes. In the underdoped sample, we observed
the existence of three energy scales --- SDW (at N\'{e}el
temperature $T_{N} \sim$ 85~K), superconductivity (at $T_{c}
\sim$ 28~K) and a normal-state phase (at $T^{\ast} \sim$ 60~K).
We observed the smooth evolution of this normal-state phase
into the SC phase, and the suppression of SDW by this
normal-state phase. We attribute this normal-state phase to the
emergence of precursor superconductivity. We incorporated the
existing ideas of phase separation and magnetic resonance mode,
and introduced new ideas of spin susceptibility
renormalization, emergence of precursor order, and the
competition between SDW and superconductivity.

Single-crystalline BKFA samples with sizes up to 10~mm x 5~mm x
0.5~mm were grown by high-temperature solution method
\cite{ChenGF08}. The grown crystals were cleaved to reveal a
fresh surface for our measurements. The values of $T_{c}$ were
confirmed by magnetization data using a Quantum Design Magnetic
Property Measurement System. No hysteresis loops in
magnetization versus field were found, ruling out the presence
of ferromagnetic impurities. The pump-probe setup was described
in Ref.~\onlinecite{Chia07}, where cross-polarized, 800~nm pump
and probe pulses, with 40~fs pulse width and 80 MHz repetition
rate, were used. The average pump power was 1-2~mW, giving a
pump fluence of $\sim$0.2 $\mu$J/cm$^{2}$. The probe intensity
was 10 times lower. Data, corrected for temperature increase of
the illuminated spot, were taken from 8~K to 140~K. The
resolution is at least 1 part in 10$^{6}$.

Figure~\ref{fig:Fig1}(b) shows the photoinduced change in
reflectance ($\Delta R/R$) of the almost optimally-doped sample
(OPT), $T_{c} \sim$ 36~K, as a function of temperature. When
the pump pulse arrives, the reflectance drops sharply and then
recovers its undisturbed value on a picosecond (ps) time scale.
We notice a very fast initial relaxation, that is
temperature-independent, of the order of the pulse width, which
we attribute to the appearance of the coherent artifact. We
will ignore it in our subsequent analysis. After the decay of
the artifact, the photoinduced change in reflectance can be
described by a single exponential decay function when the
statistical transfer of energy from quasiparticles (QP) to the
thermal reservoir is dominated by a single relaxation channel.
We extract the amplitude ($A$) and relaxation time ($\tau$) of
$\Delta R/R$ by fitting it with a single exponential decay
function, $\Delta R/R = A \exp(-t/\tau)$. The fitting
parameters $A(T)$ and $\tau (T)$ in the SC state are shown in
Figs.~\ref{fig:Fig1}(c) and ~\ref{fig:Fig1}(d).

We use the Rothwarf-Taylor (RT) model to explain our data
\cite{Rothwarf67}. This phenomenological model is used to
describe the relaxation of photoexcited SCs, where the presence
of a gap in the QP density of states (DOS) gives rise to a
bottleneck for carrier relaxation. When two QPs with energies
$\ge \Delta$ recombine ($\Delta$ = SC gap magnitude), a
high-frequency boson (HFB) with energy $\omega \ge 2\Delta$ is
created. The HFBs that remain in the excitation volume can
subsequently break additional Cooper pairs effectively
inhibiting QP recombination. Superconductivity recovery is
governed by the decay of the HFB population. From the
temperature-dependence of the amplitude $A$, one obtains the
density of thermally excited QPs $n_{T}$ via $n_{T}$$\propto$
$[A(T)/A(T$$\rightarrow$$0)]^{-1}$$-1$. Then we fit the
$n_{T}$-data to the QP density per unit cell
$n_{T} \propto \sqrt{\Delta (T) T} \exp (-\Delta (T)/T)$,
with $\Delta (0)$ as a fitting parameter. Moreover, for a
constant pump intensity, the temperature dependence of $n_{T}$
also governs the temperature-dependence of the relaxation time
$\tau$, given by
\begin{equation}
\tau^{-1}(T) = \Gamma [\delta + 2 n_{T} (T)][\Delta (T) + \alpha T
\Delta (T)^{4}], \label{eqn:tau}
\end{equation} where $\Gamma$, $\delta$ and $\alpha$ are
fitting parameters, with $\alpha$ having an upper limit of
$52/(\theta_{B}^{3}T_{min})$, $\theta_{B}$ being the
temperature of the characteristic boson, and $T_{min}$ the
minimum temperature of the experiment
\cite{Kabanov05,Demsar06,Chia06}.

Fig.~\ref{fig:Fig1}(c) shows $A(T)$ and the corresponding
$n_{T}(T)$ in its inset. Assuming that $\Delta (T)$ obeys a BCS
temperature dependence, our fits to $A(T)$ and $n_{T}(T)$ yield
$\Delta(0) = 3.0k_{B}T_{c}$, agreeing with the value obtained
from photoemission data \cite{Ding08}. We also fit $\tau (T)$
[Fig.~\ref{fig:Fig1}(d)] using Eq.~(\ref{eqn:tau}). The good
fit shows that the QP relaxation dynamics in the OPT compound
is well described by the presence of a gap in the DOS at the
Fermi level.

We now focus on underdoped BKFA ($T_{c} \sim$ 28~K) and
demonstrate that this compound exhibits a \textit{competition}
between the SC and SDW orders. The suppression of SDW order
starts at $T^{\ast} \sim$ 60~K, far above $T_{c}$. Our
subsequent analysis suggests that the Cooper pairs might
pre-form above $T_{c}$, though without phase coherence.

\begin{figure} \centering \includegraphics[width=8cm,clip]{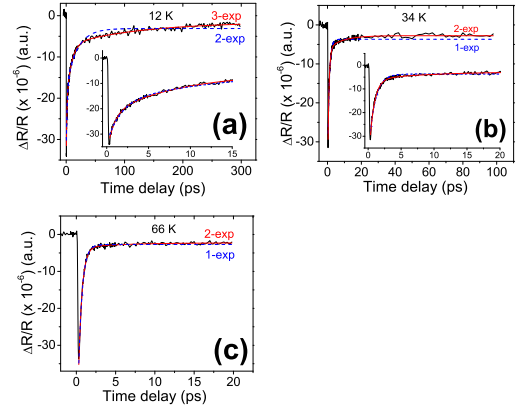}
\caption{$\Delta R/R$ in underdoped BKFA,
$T_{c} \sim$ 28~K. (a) 12 K, (b) 34 K, (c) 66 K. The solid and dashed
lines are fits to single (1-exp), double (2-exp) or triple (3-exp),
exponentials.}
\label{fig:Fig2}
\end{figure}

\begin{figure} \centering \includegraphics[width=8cm,clip]{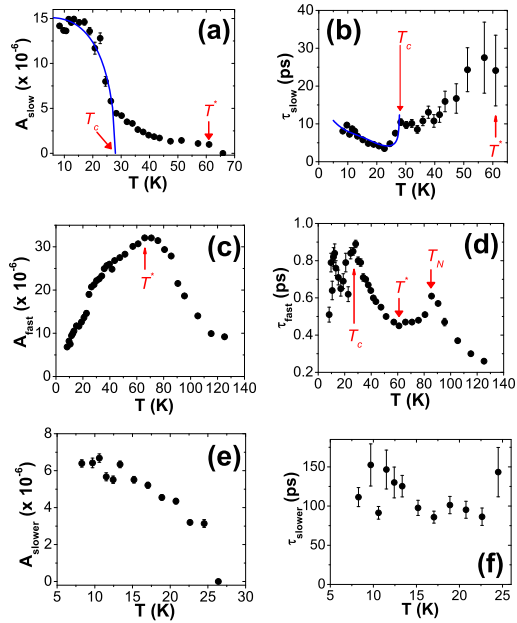}
\caption{For underdoped BKFA ($T_{c} \sim$ 28~K), values of (a) $A_{slow}$, (b) $\tau_{slow}$,
(c) $A_{fast}$, (d) $\tau_{fast}$, (e) $A_{slower}$, (f) $\tau_{slower}$,
as a function of temperature, extracted from $\Delta R/R$. In (a),
beyond $T^{\ast}$, the slow component disappears.
The solid lines in (a) and (b) are fits using the
RT model. Near $T^{\ast} \sim$ 60~K, we obtain large
error bars in $\tau_{slow}$ because the one- and two-exponential
fits of $\Delta R/R$ are of comparable quality.} \label{fig:Fig3}
\end{figure}

Figure~\ref{fig:Fig2} shows $\Delta R/R$ of underdoped BKFA at
different temperatures \cite{EPAPS}. We observe three
relaxation processes below $T_{c}$, two between $T_{c}$ and
$T^{\ast}$, and one between $T^{\ast}$ and $T_{N}$. A
three-exponential decay in the SC state was also seen in
pump-probe data of Sm(O,F)FeAs single crystals
\cite{Mertelj09}. The slow component ($\tau_{slow} \sim$
5-30~ps) corresponds to QP recombination across the SC gap, as
shown by the BCS-like temperature dependence of $A_{slow}$
below $T_{c}$ [Fig.~\ref{fig:Fig3}(a)] and by the peak in
$\tau_{slow}$ at $T_{c}$ [Fig.~\ref{fig:Fig3}(b)]. The
identical RT analysis as for the OPT sample described above
yields $\Delta (0) = 3.0k_{B}T_{c}$ [solid lines in
Fig.~\ref{fig:Fig3}(a,b)]. This shows that the opening of the
SC gap in the underdoped sample governs the QP recombination by
introducing a relaxation bottleneck.

Next, the fast relaxation component ($\tau_{fast} \lesssim$
1~ps) below $T_{N} \sim$ 85~K bears the signatures of QP
relaxation across the SDW gap: the relaxation time
$\tau_{fast}$ displays a quasi-divergence at $T_{N}$
[Fig.~\ref{fig:Fig3}(d)]. The appearance of SDW order in
underdoped BKFA, and its coexistence with superconductivity,
has also been reported in a muon spin rotation study
\cite{Aczel08}. The values of $T_{c}$ and $T_{N}$ in our
underdoped sample and that of Ref.~\onlinecite{Aczel08} are
consistent with the phase diagram in Fig.~\ref{fig:Fig1}(a).
Our measurements not only confirm the coexistence of these two
order parameters (evidenced by the existence of both the fast
and the slow relaxations below $T_{c}$), but also uncover
\textit{competition} between SDW and superconductivity, as
evidenced by the strong suppression of the SDW amplitude
($A_{fast}$) below $T_{c}$ [Fig.~\ref{fig:Fig3}(c)]. The close
proximity of the SC and SDW regions in underdoped BKFA results
in coupling between the SC and SDW order parameters, and causes
the latter to be suppressed in the SC state \cite{EPAPS}. The
suppression of the SDW order parameter in the SC state was also
observed in neutron diffraction data of the electron-doped Fe
pnictide Ba(Fe,Co)$_{2}$As$_{2}$ \cite{Pratt09}.

The sensitivity of the pump-probe technique to the presence of
SDW order is further reinforced by our study of QP relaxation
in the parent compound BaFe$_{2}$As$_{2}$ \cite{EPAPS}, with a
simultaneous SDW and first-order structural phase transition at
$T_{N} \sim$ 130~K \cite{Rotter08}. Below $T_{N}$, the
relaxation amplitude follows a BCS-like temperature dependence
down to the lowest temperatures, reflecting the behavior of the
SDW order parameter. In the vicinity of $T_{N}$, we see a
quasi-divergence in the relaxation time, which is the signature
of the opening of a gap in the DOS at the Fermi level. This,
together with our previous work in the itinerant
antiferromagnet UNiGa$_{5}$ \cite{Chia06}, shows that our
technique is sensitive to the SDW order. Data from the parent
compound thus justify our attribution of the fast relaxation in
underdoped BKFA below 85~K [Fig.~\ref{fig:Fig3}(d)] to the SDW
phase, and that the suppression of $A_{fast}$ below $T_{c}$
[Fig.~\ref{fig:Fig3}(c)] corresponds to the suppression of the
SDW order parameter.

Lastly, the slowest component ($\tau_{slower} \sim$ 100~ps)
[Fig.~\ref{fig:Fig3}(f)] is largely temperature-independent,
disappears above $T_{c}$, and corresponds to spin-lattice
relaxation. After the initial fast QP relaxation due to
electron-phonon coupling (as manifested by $\tau_{fast}$) in
the SDW region, the heated phonons then relax by transferring
their energy to the spin bath. This relaxation rate
$1/\tau_{sl} = g_{sl}/C_{spin}$, where $\tau_{sl}$ is the
spin-lattice relaxation time, $g_{sl}$ is the spin-lattice
coupling strength and $C_{spin}$ is the spin specific heat
\cite{Averitt01}. Below $T_{c}$, the 14~meV magnetic resonance
mode appears \cite{Christianson08} in the SC region. This mode
penetrates into the neighboring SDW regions and renormalizes
the imaginary part of the dynamical spin susceptibility Im$\chi
(\epsilon)$. The spin DOS in the SDW regions is given by
$N_s(\epsilon) = -(1/\pi)$Im$\chi(\epsilon)$. In the SDW
regions, above $T_{c}$, there is no renormalization of SDW
fluctuations, hence $N_s(\epsilon)$ at low energies is finite.
Below $T_{c}$, the increase in amplitude of Im$\chi (\epsilon)$
at the resonance energy removes spectral weight at lower
energies, resulting in the depression of $N_s(\epsilon)$ at low
energies. Hence $C_{spin}$ decreases, shortening $\tau_{sl}$
(=$\tau_{slower}$) to a value that is observable by our
technique \cite{EPAPS}. This scenario also explains why we do
not observe $\tau_{slower}$ in the OPT sample --- there are no
SDW regions to begin with, hence no SDW fluctuations for the
magnetic resonance mode to renormalize.

In addition to establishing the competition between SDW and
superconductivity, the data in Fig.~\ref{fig:Fig3}(a-d) carry
evidence of a \textit{precursor order} (PO) that appears at
$T^{\ast} \sim$ 60~K in the normal state of underdoped BKFA. In
Fig.~\ref{fig:Fig3}(a), $A_{slow}$ exhibits a well-defined tail
that survives well above $T_{c}$, and disappears above
$T^{\ast}$. Compare this to $A_{slow}$ of the OPT sample
[Fig.~\ref{fig:Fig1}(c)], where no such tail exists. This
suggests that a kind of precursor superconductivity has already
existed between $T_{c}$ and $T^{\ast}$. A tail in the
relaxation amplitude, attributed to the pseudogap, was also
seen in underdoped Y-123 \cite{Kabanov99}. Moreover, in
addition to a quasi-divergence of $\tau_{slow}$ at $T_{c}$
[Fig.~\ref{fig:Fig3}(b)], indicative of the opening up of a SC
gap, $\tau_{slow}$ continues to increase above $T_{c}$ and
peaks at $T^{\ast}$, showing that, at $T^{\ast}$, another QP
gap opens up at the Fermi level. Compare this to a typical SC,
where $\tau_{slow}$ plunges to the metallic value of
$\sim$0.5~ps immediately after $T_{c}$, and remains almost
temperature-independent above $T_{c}$. We have also taken data
on another underdoped BKFA with a similar $T_{c}$, and obtained
a similar $T^{\ast}$ of 55~K.

Further evidence for precursor pairing at $T^{\ast}$ comes from
the temperature dependence of $A_{fast}$ --- the SDW amplitude.
In the cuprate HTSCs, the proximity of \textit{d}-wave
superconductivity to antiferromagnetism is simply assumed as an
experimental fact. However, from a microscopic point of view,
\textit{d}-wave superconductivity in the cuprates turns out to
be the winner of the competition between these two orders
\cite{Lee06} --- this statement may also hold for the extended
$s_{\pm}$ pairing that the pnictide SCs are thought to have,
where the gaps at the hole and electron pockets are of
\textit{opposite} signs to each other \cite{Mazin08}. In our
underdoped sample, if a PO develops at $T^{\ast}$, the SC
fluctuations associated with the PO will start to ``win over'',
ie. suppress, SDW even in the normal state. This would explain
the suppression of SDW below $T^{\ast}$ in our underdoped
sample [Fig.~\ref{fig:Fig3}(c)]. The rather broad peak in
$\tau_{slow}$ at $T^{\ast}$ might indicate the presence of
disorder. It implies that, though the PO develops at
$T^{\ast}$, disorder may cause the QP excitations to be
partially gapless.

Due to partial Fermi surface nesting, the formation of SDW only
induces a partial gap opening on the Fermi surface
\cite{Dong08}. Thus, in the SDW islands, the fast relaxation
rate is the sum of QP relaxations across the gapped and
ungapped regions of the SDW Fermi surface: $1/\tau_{fast} =
1/\tau_{fast}^{gapped} + 1/\tau_{fast}^{gapless}$. Below
$T_{c}$, the rise of the magnetic resonance mode in the
neighboring SC regions renormalizes the SDW fluctuations in the
SDW islands, making $\tau_{fast}^{gapless}$, and hence
$\tau_{fast}$, smaller than the corresponding $\tau_{fast}$ in
the normal state. This explains the downturn of $\tau_{fast}$
in the SC state.

Recent photoemission data offered evidence of precursor pairing
in the iron pnictide SCs, such as in La(O,F)FeAs
\cite{Sato08,Ishida08} and Sm(O,F)FeAs \cite{Liu08}. The Nernst
effect in La(O,F)FeAs also suggested the presence of a
``precursor state'' between $T_{c}$ and 50 K in which magnetic
fluctuations are strongly suppressed \cite{Zhu08}. A recent
pump-probe study of Sm(O,F)FeAs gave evidence of a
pseudogap-like feature with an onset around 200~K
\cite{Mertelj09}. In our case, the PO is associated with an
intermediate energy scale $T^{\ast}$ between magnetism and
superconductivity. Our PO does not compete with
superconductivity, but competes with the SDW order. The PO that
sets in at $T^{\ast}$ seems to be intimately related to
superconductivity, as its signature becomes the signature of
superconductivity below $T_{c}$. Therefore, we suggest that the
PO may be a precursor of the SC order, much like the Cooper
pairing without phase coherence that precedes macroscopic
superconductivity in cuprate HTSCs. The detailed nature of the
PO, whether it is due to phase fluctuations \cite{Lee06} or
their interplay with disorder, remains an open scientific
question. Finally, note that the ratio $T^{\ast}/T_{c} >2 $ has
also been observed in underdoped cuprates \cite{Fischer07}, and
various theories \cite{Anderson87,Levin05} have attempted to
account for it.

In conclusion, our results on underdoped (Ba,K)Fe$_{2}$As$_{2}$
suggest the existence of precursor superconductivity above
$T_{c}$ that suppresses antiferromagnetism. We also offer
evidence of the renormalization of SDW fluctuations by the
magnetic resonance mode, in the framework of mesoscopic phase
separation and partial SDW Fermi surface nesting.

This work was supported by the LANL LDRD program, the Singapore
Ministry of Education AcRF Tier 1 (RG41/07) and Tier 2
(ARC23/08), the National Science Foundation of China, the 973
project and the National Basic Research Program of China
(Contract No. 2009CB929104), the Chinese Academy of Sciences,
PCSIRT of the Ministry of Education of China (Contract No.
IRT0754), MEXT-CT-2006-039047 and the National Research
Foundation of Singapore.

\bibliographystyle{prsty}
\bibliography{FeAs}
\bigskip

\end{document}